\input harvmac
\Title{\vbox{\baselineskip12pt\hbox{CERN-TH/96-179}}} {\vbox{\centerline
{Topological Defects in Gauge} \vskip2pt\centerline{Theories of Open
$p$-branes}}}
 \centerline{M. C. Diamantini} \medskip
\centerline{CERN Theory Division, CH-1211 Geneva 23,
Switzerland}
\vskip .3in
\centerline{Abstract}
We study phase transitions induced by topological defects in
compact Abelian gauge theories of {\it open} $p$-branes in (d+1)
space-time dimensions.
Starting from a massive antisymmetric tensor theory for the open
$p$-branes we show how the condensation of
topological defects can lead to a decoupled phase with a massless
tensor coupled to closed $(p-1)$-branes and a massive tensor coupled to
open $(p+1)$-branes. We also consider the case, relevant in string
theory, in which the boundaries of the $p$-branes are constrained to live
on a Dirichlet $n$-brane.

\Date{$\eqalign{&{\rm CERN-TH}/96-179\cr &{\rm July}\ 1996\cr}$}
\vfill
\eject
Antisymmetric tensors of rank $(p+1)$\ ($(p+1)$-forms) have been widely
studied in  recent years  \ref\kal{V.I.
Ogievetsky and I.V. Polubarinov, Sov. J. Nucl. Phys. 4 (1967) 156;
M. Kalb and P. Ramond, Phys. Rev. D9 (1974) 2273.}, \ref\nam{Y.
Nambu, Phys. Rep. 23 (1976) 250; E. Cremmer and J. Scherk, Nucl. Phys.
B72  (1974) 117; D.Z. Freedman and P.K. Townsend, Nucl. Phys. B177
(1981) 282.}, \ref\tei{ R.I. Nepomechie,
Phys. Rev. D31 (1984) 1921; C. Teitelboim, Phys. Lett. B167 (1986)
63; C. Teitelboim, Phys. Lett. B167 (1986) 69.}, \ref\orl{R. Savit, 
Phys. Rev. Lett. 39 (1977) 55; P. Orland,
Nucl. Phys. B205[FS5] (1982) 107.For a review see: R. Savit,
Rev. Mod. Phys. 52 (1980) 453.},
\ref\pin{F. Quevedo and C.A. Trugenberger, CERN-TH/96-109,
hep-th/9604196.}. They arise naturally in constructing Abelian $U(1)$\
gauge theories of elementary extended  objects (strings, membranes,...
): a $(p+1)$\ antisymmetric tensor couples to elementary $p$-branes in
the same way as the vector potential one-form in Maxwell theory couples
to elementary point-particles (0-branes). Antisymmetric tensors appear
in the effective field theories for the low-energy dynamics of strings
and in supersymmetric theories \ref\gre{M. Green, J. Schwarz and E.
Witten, ``Superstring Theory'', Cambridge University Press, Cambridge
(1987).}.

As in the familiar case of QED \ref\pol{For a review see: A.M. Polyakov,
``Gauge Fields and Strings'', Harwood Academic Publishers, Chur
(1987).}, the compactness of the gauge symmetry implies the existence of
topological defects. These play a crucial role in gauge theories and
can lead to drastic modifications of the perturbative infrared
behaviour of a theory. Both $U(1)$ and $R$ lead to the same
perturbative expansion, but in the former case the condensation of
topological defects can generate a mass for the gauge fields and cause
confinement. 

The topological defects are $({\rm d}-p-2)$-dimensional extended objects
representing the world-hypervolumes of $({\rm d}-p-3)$-branes
(instantons are ($-1$)-branes). The possible phase diagram depends on
the space-time dimension and on the dimension of the topological
defects. This can be easily seen if we think of compact QED: in 3
dimensions the topological excitations are ($-1$)-branes (instantons);
these are in a plasma phase for all values of the
coupling constant, leading to electric charge confinement
\pol. In 4 dimensions the topological defects are
0-branes (monopoles) and they condense only above a critical value of
the coupling constant.

For closed or infinitely extended $p$-branes the action that describes
the model can be written as 

\eqn\cpb{\eqalign{ S &= \int {(-1)^{p+1}\over g^2} dB_{p+1} \wedge
*dB_{p+1} + j B_{p+1} \wedge * J_{p+1}\cr 
&= \int {(-1)^{p+1}\over g^2} dB_{p+1} \wedge
*dB_{p+1} + j\int_{\rm branes} B_{p+1} \ ,\cr }}
where $B_{p+1}$\ is the antisymmetric tensor field, $g$ is
a dimensionless coupling constant, $j$ has dimension $[m^{d-2p-3
\over 2}]$,  and $J_{p+1}$\ is a conserved (tensor) current
of closed or infinitely extended $p$-branes :

\eqn\ccp{J^{\mu_1...\mu_{p+1}} (x) = \int 
\delta^{d+1}(x - y(\sigma))\ dy^{\mu_1} \wedge...\wedge dy^{\mu_{p+1}}
\ ,}
where $y(\sigma)$\ are the coordinates of the world-volumes of the
$p$-branes.
The action \cpb\ is the same action as the one that describes the
coupling  of a Dirichlet $p$-brane to a $(p+1)$ form \ref\poc{J.
Polchinski, S. Chaudhuri and C.V. Johnson, ``Notes on D-branes'',
hep-th/9602052 and references therein.}.
We will consider the current $J_{p+1}$\ as an external probe and ignore 
the action governing its dynamics.

For $p$ = 1 and d+1 = 4 \cpb\ describes the classical interstring
action for closed strings \kal.
The effect of the presence of topological defects in \cpb\ has been
studied in \orl, \ref\sjr{S.J. Rey, Phys. Rev. D10
(1989) 3396.}\ using a lattice regularization in
Euclidean space-time.
By means of a duality transformation on the lattice, \cpb\ can be
transformed into a theory formulated in terms of integer fields and
this formulation can be used to describe the interactions of
topological defects.
What has been found \orl\ is that the theory is disordered
(condensation of the topological defects) for all couplings if d =
$p$+1, while a critical non-zero coupling exists for d = $p$+2.
When the system is disordered, the effect of the condensation of the
topological defects is the {\it confinement} of closed $p$-branes.

In this paper we are interested in considering the effects of
topological excitations in a $U(1)$ gauge theory for {\it open}
$p$-branes.
The theory we consider is:

\eqn\opb{\eqalign{ S = &\int {(-1)^{p+1}\over g^2} dB_{p+1} \wedge
*dB_{p+1} + (-1)^p \left( m B_{p+1} +{1 \over e} dA_p \right) 
\wedge * \left( m B_{p+1} +{1 \over e} dA_p \right) \cr 
&+ j \left ( B_{p+1} + {1\over me} dA_p \right)
 \wedge * J_{p+1}\ , \cr }}
where $A_p$\ is an antisymmetric tensor of rank $p$, $e$\ is a
dimensionless coupling constant, $m$ is a mass parameter, and $S$ is
invariant under the combined gauge symmetry:

\eqn\cgs{\eqalign{ B_{p+1} \quad &\rightarrow B_{p+1} + d\Lambda_p\ ,
\cr A_p \quad &\rightarrow A_p - me\Lambda_p\ . \cr }}
The term $\left ( B_{p+1} + {1\over me} dA_p \right)$ is itself 
gauge-invariant; for this reason we do not need the current $J_{p+1}$ to be
conserved, and we can couple the theory to open $p$-branes.

Again, for $p$ = 1 and d+1 = 4, \opb\ describes the classical 
interstring action for open strings \kal, \sjr. 
In this case $dA$ is included to ensure gauge invariance at the
end-points of the open string.
More generally, if we look at \cpb\ we see that the coupling is
invariant only if the world-hypervolume of the $p$-brane has no
boundary. The antisymmetric tensor $A_p$ couples to the boundary and
restores gauge invariance.

Also interesting to study is a slight modification of \opb, namely:

\eqn\dba{\eqalign{ S = &\int_M {(-1)^{p+1}\over g^2} dB_{p+1}
\wedge *dB_{p+1} + \int_\Sigma (-1)^p \left( m B_{p+1} +{1 \over e}
dA_p \right)  \wedge * \left( m B_{p+1} +{1 \over e} dA_p \right) \cr 
&+ \int_M j B_{p+1}  \wedge * J_{p+1} + \int_\Sigma j {1\over me} A_p
 \wedge * \delta J_{p+1} \ . \cr }}
Here $M$ indicates the full (d+1)-dimensional space-time, while
$\Sigma$ is an infinitely extended $n$-dimensional hypersurface in our
space-time ($({\rm d}+1) \ge n$) on which the boundary of the open $p$-brane
is constrained to live. In this case $m$ has dimension $[{\rm
mass}^{n-d+1\over 2}]$. We recover \opb\ when (d+1) = $n$. The
action \dba\ is a generalization, in flat space-time, of the effective
action \ref\wit{E. Witten, IASSNS-HEP-95-83,
 hep-th/9510135.}\ for Type II superstrings in 10
dimensions in presence of a D-brane (a D-string in the particular
case $n$ = 2):  

\eqn\ewa{S = \int_M d^{10} x \sqrt{g} {1\over 2\lambda^2} dB_2
\wedge * dB_2 + \int_\Sigma d^2 x \sqrt{g_\Sigma} {1\over 2\lambda}
\left(B_2 + dA \right) \wedge * \left(B_2 + dA \right) \ ,}
where $\lambda$ is the string coupling constant and $g$ and
$g_\Sigma$ are the metrics on $M$ and $\Sigma$.
D-branes in Type II superstrings are BPS states that carry RR
charges \poc. The necessity of objects that carry RR charges 
comes from string dualities \ref\hul{C.M. Hull and P.K. Townsend,
Nucl. Phys. B438 (1995) 109; E. Witten, Nucl. Phys. B438 (1995) 85;
A. Strominger, hep-th/9504090.}.
The end-points of open strings are constrained to live on the
brane. This is why the part of the action that contains the $U(1)$
gauge field $A$, which couples to the boundaries of the open string,
is defined on the D-brane world-sheet.
In our simplified model, the D-brane is considered as a submanifold
without dynamics on which the boundaries of the open $p$-branes are
constrained to live. This allows us to consider the
the generic case of open $p$-branes ending on $n$-branes with the only
constraint $(d+1) \ge n \ge p$, without taking into account consistency
coming from string theory \poc\ \ref\dal{M.Douglas,
preprint RU-95-92, hep-th/9512077.} \ref\str{A. Strominger, preprint
hep-th/9512059.} \ref\tow{P.K. Townsend, preprint
DAMTP-R-95-59, hep-th/9512062.}.
Note also that fermion zero-modes in supersymmetric theories may
crucially affect the dynamics of topological excitations \ref\nsj{We
thank S.J. Rey for pointing this out.}.

We will start with the analysis of the first model \opb\ and then come
back to \dba.  
In \opb\ the tensor of higher-rank possesses a bare mass term.
In fact, up to a gauge transformation, we can rewrite \opb\ as a
generalized version of the Proca Lagrangian for an
antisymmetric tensor field $\bar B_{p+1} = B_{p+1} + {1 \over me}\
dA_p$:

\eqn\sla{\eqalign{ S = &\int {(-1)^{p+1}\over g^2} d \bar B_{p+1} \wedge
* d \bar B_{p+1} + (-1)^p m^2 \bar B_{p+1}\wedge * \bar B_{p+1} \cr
  &+ j \bar B_{p+1} \wedge * J_{p+1}\ .\cr }}
In this formulation the higher-rank tensor $B_{p+1}$ has ``eaten'' the
tensor $A_p$, which is a generalization of the familiar
Higgs-St\"uckelberg mechanism for vector fields. Correspondingly, \sla\
describes a generalized ``Higgs'' phase \sjr.
When (d+1) = 4 and $p$ = 1 the integral over $B_2$ defines the action
for the {\it confining string} $J_2$ \pin, \ref\spo{A.M. Polyakov,
hep-th/9607049.}. What we
want to do now is to take into account the presence of topological
defects in the theory with compact symmetry, and study what the
effects of their possible  condensation are.

In general, when studying the effect of condensation of topological
excitations, one starts with a compact theory in the Coulomb phase. The
condensation of topological defects, when possible,
disorders the system and generates a mass for the gauge field.

We know from the St\"uckelberg mechanism \ref\iez{For a review  see: C.
Itzykson and J.B. Zuber, ``Quantum Field Theory'', McGraw-Hill, New
York (1980).} that the limit in which the mass goes to zero in \sla\ is
singular, because it introduces an additional gauge symmetry, not
present in the original theory. The presence of a massless pole in the
propagator for the $B_{p+1}$ tensor, in the limit $m \to 0$, can be
seen by adding a gauge-fixing term; in order to recover a
gauge-invariant theory in the limiting case, however, it is necessary
that the tensor field be coupled to a conserved current, and in our
case $J_{p+1}$ is not (current of open $p$-branes). As we will see, 
there is also in this case the possibility of the appearance of a
gauge-invariant massless pole in the propagator of the $B_{p+1}$ tensor.
What leads to the zero-mass limit in our theory is the condensation of
topological excitations; apparently, in this case, these work in the
opposite direction with respect to the general case by cancelling the
mass for the gauge field $B_{p+1}$. As we will show, after the
condensation of the topological defects, the $B_{p+1}$ tensor couples
automatically only to the transverse part of $J_{p+1}$; the
longitudinal part decouples and describes a generalized Coulomb gas.
The transverse part represents a conserved current and the theory is
gauge-invariant.

The topological defects can be treated explicitly by rewriting \opb\
as:

\eqn\opd{\eqalign{ S = &\int {(-1)^{p+1}\over g^2} \left( dB_{p+1}
+ q t_{p+2} \right) \wedge * \left( dB_{p+1} +q t_{p+2} \right) +\cr 
&+ (-1)^p \left( m B_{p+1} +{1 \over e} dA_p + {s \over e} t_{p+1}
\right)  \wedge * \left( m B_{p+1} +{1 \over e} dA_p + {s\over
e} t_{p+1} \right) +\cr  
&+ j \left( B_{p+1} + {1\over me} A_p \right) \wedge * J_{p+1}\ .\cr
}}
Here $t_{p+2}$ and $t_{p+1}$ are singular forms representing 
topological defects of charge $q$ and $s$ \tei, \pin\ associated with
the two compact gauge fields $B_{p+1}$ and $A_p$:

\eqn\tdf{ *t_{r} = G_{d-r+1}\ ,\quad 
G^{\mu_1...\mu_{d-r+1}}_{d-r+1} =  \int \delta^{d+1}\left( x -
\tilde y(\tilde \sigma) \right) d \tilde y^{\mu_1} \wedge ...\wedge
d \tilde y^{\mu_{d-r+1}} \ .}
Here $\tilde y^{\mu} (\tilde \sigma^{\nu_1},...,\sigma^{\nu_{d-r+1}})$
is an open hypervolume describing the generalization to
higher-dimensional topological defects of the Dirac string. The
boundary of this hypervolume describes the world-hypersurface of the
topological defects. 

Because the forms $t_r$ are singular, their treatment requires a
proper regularization. For vector fields we can think of  Abelian
compact gauge models as the low-energy effective field theory due
to spontaneous symmetry breaking of a non-Abelian gauge
group \pol. In this case the mass acquired by the broken gauge fields
becomes the cut-off for the low energy theory. 
Another possibility is to use a lattice regularization, which is the
route  we will eventually choose.

By integrating over the tensor field $A_p$ we obtain an effective action
for the higher rank-tensor $B_{p+1}$. When the topological defects of
$A_p$ are dilute, this effective action still possesses a massive pole,
while in the case in which these topological defects are in a dense
phase, the effective action we get is (ignoring for the
moment the coupling with the current of open $p$-branes):

\eqn\ecb{S_{\rm eff} = \int {(-1)^{p+1}\over g^2} \left( dB_{p+1}
+ q t_{p+2} \right) \wedge * \left( dB_{p+1} + q t_{p+2} \right) \ .}
This is nothing else than the compact version of \cpb: the
mass term for $B_{p+1}$ is no longer present.
If we count the degrees of freedom in \sla\ we obtain 
$\pmatrix{d \cr p+1 \cr}$; this expression can be written as:

\eqn\dof{\pmatrix{d \cr p+1 \cr} = \pmatrix{d-1 \cr p+1 \cr} +
\pmatrix{d-1 \cr p \cr} \ .}
Equation \dof\ is the sum of the degrees of freedom of the two massless
free theories for the two antisymmetric tensors $B_{p+1}$ and $A_p$: in
\sla\ the tensor $B_{p+1}$ ``eats'' $A_p$ and becomes massive through 
the Higgs-St\"uckelberg mechanism. The condensation of the topological
defects of $A_p$ prevents this mechanism.

In the coupling with the $(p+1)$-current of open $p$-branes, $j \int
\left ( B_{p+1} + {1\over me} dA_p \right)\wedge *J_{p+1}$, we use the
Hodge decomposition for the form $J_{p+1}$:

$$J_{p+1} = d \beta_p + \delta \alpha_{p+2} \ ,$$
where $\delta$ is the adjoint exterior derivative.
The properties of the inner product between forms and the fact that
$\delta^2 = 0$ tell us that $dA_p$ couples only to $d\beta_p$:
$\left( dA_p , d\beta_p \right) = \left( A_p , \delta d\beta_p
\right)$, where $\delta d\beta_p$ is the hypersurface which represents
the boundary of the open hypervolume $J_{p+1}$ and (,) denotes the
inner product.
In the effective action \ecb, as we will show, the antisymmetric
tensor $B_{p+1}$ couples only to the transverse part of $J_{p+1}$,
$\delta \alpha_{p+2}$, while the boundaries $\delta d\beta_p$
decouple and form a generalized Coulomb gas. 
This can be represented as a Gaussian integral over a massless $p$-form
$A_p$ coupling to $\delta d\beta_p$.

What happens if also the topological defects of $B_{p+1}$ condense?
This can be studied analogously by computing the effective action for
$J_{p+1}^T$ obtained by integrating over $B_{p+1}$.In the dense phase
we obtain a kernel, which can be represented as a Gaussian integral
over a new, massive, $(p+2)$-form $C_{p+2}$ coupling to $\alpha_{p+2}$.
In this phase we thus obtain two decoupled gauge theories for open
$(p-1)$-branes and closed $(p+1)$-branes. These describe 
$\pmatrix{d \cr p+2 \cr}$ massive and $\pmatrix{d-1 \cr p \cr}$
massless degrees of freedom, respectively.

In order to properly derive this mechanism in a specific  case of \opb,
we consider a Euclidean lattice regularization for the case $p$ = 1,
d+1 = 4. In this case $J_{p+1}=J_2$ describes the world-sheet of an open
string, and its boundary is the closed world-line described by the
end-points of the open string.

The continuum non-compact Euclidean action is:

\eqn\cna{S = \int d^4x \left( {1\over 2g^2} F_\mu F_\mu + {1\over
4e^2} f_{\mu \nu} f_{\mu \nu} + {\tilde m^2 \over 4} B_{\mu \nu}
B_{\mu \nu} + {\tilde m \over 2e} B_{\mu \nu} f_{\mu \nu} \right) \
,}
where $B_{\mu \nu}$ is the antisymmetric Kalb-Ramond tensor, $F_\mu
= {1\over 6} \epsilon_{\mu \nu \alpha \beta} F_{\nu \alpha \beta}$ is
the dual of the Kalb-Ramond field strength, $F_{\nu \alpha \beta} =
\partial_{[\nu} B_{\alpha \beta]}$, $f_{\mu \nu} = \partial_\mu
A_\nu - \partial_\nu A_\mu$ is the Maxwell field strength, and we
have renamed the mass parameter as $\tilde m$.
Equation \cna\ is invariant under the combined gauge transformation

\eqn\ngt{\eqalign{ B_{\mu \nu} \quad &\rightarrow B_{\mu \nu} +
\partial_\mu \Lambda_\nu - \partial_\nu \Lambda_\mu \cr A_\mu \quad
 &\rightarrow A_\mu - \tilde m e\Lambda_\mu\ . \cr }}

The lattice we consider is a hypercubic lattice in four Euclidean
dimensions, whose sites are denoted by $x$, and lattice spacing $l$.
The gauge fields, which are angular variables for 
the compact theory on the lattice, are associated with the
links $(x, \mu)$ between the sites $x$ and $(x+\hat \mu)$, where $\hat
\mu$ denotes a unit vector in the direction $\mu$ on the lattice.
On the lattice, we define the following forward and backward 
derivatives and shift operators: 

\eqn\der{\eqalign{d_\mu f({\bf x}) &\equiv {f({\bf x} + \hat \mu l) -
f({\bf x}) \over l}\ , \qquad S_\mu f({\bf x)} \equiv f({\bf x} +
\hat \mu l) \ , \cr
\hat d_\mu f({\bf x}) &\equiv {f({\bf x}) - f({\bf x} - \hat \mu l)
\over l} \ , \qquad \hat S_\mu f({\bf x}) \equiv f({\bf x} - \hat \mu
l) \ . \cr}}
Summation by parts interchanges the two derivatives, with a minus sign,
and the two shift operators.
We also introduce the three-index lattice operators \ref\cri{M.C.
Diamantini, P. Sodano and C.A. Trugenberger, preprint
CERN-TH/95-294, to appear in Nucl. Phys. B.}:

\eqn\ddk{K_{\mu \nu \alpha} = S_\mu \epsilon_{\mu \rho \nu \alpha}
d_\rho \ , \quad \hat K_{\mu \nu \alpha} =  \epsilon_{\mu \nu
 \rho \alpha} \hat d_\rho \hat S_\alpha \  .}
These operators are gauge-invariant in the sense
that:

\eqn\pdk{\eqalign{K_{\mu \nu \alpha} d_\alpha &= K_{\mu \nu \alpha}
d_\nu = \hat d_\mu K_{\mu \nu \alpha} = 0 \ , \cr
\hat K_{\mu \nu \alpha} d_\alpha &= \hat d_\mu \hat K_{\mu \nu
\alpha} = \hat d_\nu \hat K_{\mu \nu \alpha} = 0\ .\cr}}
Moreover they satisfy the equations:

\eqn\ker{\eqalign{&\hat K_{\mu \nu \alpha} K_{\alpha \lambda \omega} =
K_{\mu \nu \alpha} \hat K_{\alpha \lambda \omega} = O_{\mu \nu
\lambda \omega } =\cr
&= - \left( \delta_{\mu \lambda} \delta_{\nu \omega} - 
\delta_{\mu \omega} \delta_{\nu \lambda}\right) \nabla^2 + \left( 
\delta_{\mu \lambda} d_\nu \hat d_\omega - \delta_{\nu \lambda}
d_\mu \hat d_\omega \right) + \left( 
\delta_{\nu \omega} d_\mu \hat d_\lambda - \delta_{\mu \omega}
d_\nu \hat d_\lambda \right) \ ,\cr
&\hat K_{\mu \omega \alpha} K_{\omega \alpha \nu } =
K_{\mu \omega \alpha} \hat K_{\omega \alpha \nu } =  2 M_{\mu \nu} = - 
2 \left(  \delta_{\mu \nu} \nabla^2 - d_\mu \hat d_\nu \right) \ . \cr}}
The expressions $O_{\mu \nu \lambda \omega }$ and $M_{\mu \nu}$ are 
lattice versions of the Kalb-Ramond and Maxwell kernels, respectively,
and $\nabla^2 = d_\mu \hat d_\mu = \hat d_\mu d_\mu$ is the lattice
Laplacian. 

The formulation of the theory on the lattice allows a quantitative
analysis of the phase diagram and the condensation conditions for the
topological excitations. We consider a partition function of the
Villain \ref\kle{For a review see: H. Kleinert, ``Gauge Fields in
Condensed Matter'', World Scientific, Singapore (1989).}\ type:

\eqn\lam{\eqalign{Z(J) &= \sum _{{\{\tilde b_{\mu }\}, \{\tilde a_{\mu
\nu }\} }\atop {  \{ m_{\mu \nu }\} }} \int _{-\pi \over
l}^{\pi \over l}
 {\cal D} A_{\mu } {\cal D} B_{\mu \nu } \ {\rm exp}(-S) \ ,\cr
S &= \sum_{{\bf x}, \mu }\  {l^4\over 2g^2} \left( F_{\mu } +{\pi
\over l^2} \tilde b_{\mu } \right) ^2 + {l^4 \over 4} \tilde m^2
 \left( B_{\mu \nu} +{2\pi \over l} m_{\mu \nu } \right)^2 
 +{l^4\over 4e^2} 
\left( f_{\mu \nu} + {2\pi \over l^2} \tilde a_{\mu \nu} \right)^2
\cr
&+ {l^4 \over 2e} \tilde m \left( B_{\mu \nu} +{2\pi \over l} m_{\mu \nu }
 \right) \left( f_{\mu \nu} + {2\pi \over l^2} \tilde a_{\mu \nu} 
\right)^2 - ilj \left( B_{\mu \nu} + { 1\over \tilde m e} f_{\mu \nu}
\right) J_{\mu \nu} \ .\cr}}
The dual of $F_{\mu \nu \alpha}$ is expressed in terms of the $K_{\mu
\nu \alpha}$ operator as: $F_\mu = {1\over 2} K_{\mu \nu \alpha}
B_{\nu \alpha}$. The integer fields $\tilde a_{\mu \nu},\ \tilde
b_\mu$ and $m_{\mu \nu}$  ensure the periodicity under the
transformations

\eqn\shi{A_\mu \rightarrow A_\mu + {2\pi \over l} n_\mu\quad {\rm
and}\quad   B_{\mu \nu} \rightarrow B_{\mu \nu} + {2\pi \over l} n_{\mu
\nu}\ ,} with $n_\mu$ and $n_{\mu \nu} \in {\cal Z}$.
The action in \lam\ possesses also the gauge symmetry \ngt\ with
derivatives substituted by lattice derivatives $d$.

The last term in \lam, the coupling with the world-sheet of an open
string, plays the role of the (non-local) order parameter, the Wilson
``surface'' $W_S$, for the phase transitions in the theory.
$J_{\mu \nu}$ vanishes everywhere but on the plaquettes of the Wilson
surface, where it takes the value 1. The integer $j$ represents the
strength of the coupling. The most general case in which the last term
in \lam\ is invariant under \shi, is when: \eqn\qco{{1 \over l \tilde m
e } = {q\over p}\ {\rm and}\ j = \tilde j p}
with $p$, $q$ and $\tilde j\ \in {\cal Z}$ and $(p,q)$ coprime.

Now we decompose the integers $\tilde a_{\mu \nu}$ and $\tilde b_\mu$ as
\eqn\sho{\eqalign{\tilde a_{\mu \nu} &= l (d_\mu \lambda_\nu - d_\nu
\lambda_\mu) + a_{\mu \nu} \ , \cr
\tilde b_\mu &= l K_{\mu \nu \alpha} m_{\nu \alpha} + b_\mu\
,\cr}}
and  change variables $B_{\mu \nu} \rightarrow B_{\mu \nu} + {2 \pi
\over l} m_{\mu \nu}$, $A_\mu \rightarrow A_\mu + {2 \pi \over
l} \lambda_\mu$.
The sum over $m_{\mu \nu}$ and $\lambda_\mu$ has the effect of
shifting the integrals in the partition function from $[-\pi/l, \pi /l]$
to $(-\infty, +\infty)$. We can now perform the Gaussian integrals over
the gauge fields to compute the expectation value 
$\langle W_S \rangle = Z(J)/Z$ of the Wilson surface:

\eqn\top{\eqalign{\langle W_S \rangle &= 
{1\over Z_{\rm top} } \sum_{{\{c_{\mu }\}}
\atop {\{b_{\mu }\}}} {\rm exp } \left( - S_{\rm top} - W_{\rm top} -
W_0 \right) \cr  S_{\rm top} &= \sum_{{\bf
x}, \mu } \ {\pi^2\over 2g^2} \  b_{\mu } {{m^2 \delta _{\mu \nu
} -d_{\mu }\hat d_{\nu }} \over {m^2-\nabla ^2}}b_{\nu } 
+{\pi^2 \over 2e^2 l^2}\  c_{\mu }{1 \over {m^2-\nabla ^2}}
c_{\mu } - {\pi^2 m \over l g e}\  b_{\mu }
{1\over {m^2-\nabla ^2}}
c_{\mu }\ ,\cr
W_0 &= \sum_{{\bf x}, \mu } {g^2  j^2\over l^2}\  J_{\mu \nu}{1 \over
{m^2-\nabla ^2}} J_{\mu \nu } + {2 g^2  j^2\over m^2 l^2}\  \hat
d_\mu J_{\mu \nu}{1 \over {m^2-\nabla ^2}} \hat d_\rho J_{\rho \nu }
\ ,\cr
W_{\rm top} &= \sum_{{\bf x}, \mu } {i \pi j\over l}\  b_\mu
{\hat K_{\mu \nu \alpha} \over {m^2-\nabla ^2}} J_{ \nu \alpha } + {
i \pi j g m\over e l^2}\ c_\mu {K_{\mu \nu \alpha} \over {m^2-\nabla
^2}} J_{\nu \alpha } \ .\cr}}
Here $c_\mu = l K_{\mu \nu \alpha} a_{\nu \alpha}$ is the physical
integer degree of freedom that describes the topological defects
associated with the integer part of the Maxwell gauge field. It
describes closed (or infinitely long) strings of magnetic charge: $\hat
d_\mu c_\mu = 0$. Note that, due to this constraint and to the gauge
invariance of $\lambda_\mu$ in \sho, the set $\{c_\mu,\lambda_\mu\}$
describes 6 integer degrees of freedom, which are equivalent to the
original antisymmetric integers $\tilde a_{\mu \nu}$.
The strings $b_\mu$ come, instead, from the integer
part of $B_{\mu \nu}$; they can be either open or closed. The parameter
$m = g \tilde m$.

In order to establish the phase structure of the
model, we need to analyse the conditions
for the condensation of the topological excitations. To this end we
shall use the same free-energy arguments as those adopted in the
analysis of  related (3+1)-dimensional models \ref\eli{S. Elitzur, R.
Pearson and J. Shigemitsu, Phys. Rev. D19 (1979) 3698; D. Horn, \quad
M. Weinstein and S. Yankielowicz, Phys. Rev D19 (1979) 3715; A. Guth, A.
Ukawa and P. Windey, Phys. Rev. D21 (1980) 1013.} \ref\cra{J.L.
Cardy and E. Rabinovici, Nucl. Phys. B205 [FS5] (1982) 1; J.L.
Cardy, Nucl. Phys. B205 [FS5] (1982) 17; A. Shapere and F. Wilczek,
Nucl. Phys. B320 (1989) 669.}, in which 
the condition for the condensation of strings is established by
comparing their self-energy and their entropy.

The entropy of a string of length $L=lN$  can be 
estimated using the theory of random walks \ref\ied{For a review see:
C. Itzykson and J.M. Drouffe, ``Statistical Field Theory'',
Cambridge University Press, Cambridge (1989).} as $\gamma N$: the
parameter $\gamma $ is given roughly by $\gamma ={\rm ln} 7$, since at
each step the string can choose between 7 different directions. In a
dilute instanton approximation, in which all values $c_{\mu }, b_{\mu }
\ge 2$ are neglected, it can be proved that the correct value of
$\gamma $ is the same for open and closed strings \ref\eis{M.B.
Einhorn and R. Savit, Phys. Rev. D19 (1979) 1198.}. We will neglect
all subdominant functions of $N$, such as a ${\rm ln} N$ correction to
the entropy. 
The free energy of a string of length $L=lN$ carrying quantum
numbers $c$ and $b$ is thus essentially

\eqn\pse{ F= \left \{ {\pi^2 \over 2 g^2} G(ml) \left( b - {q\over p} c
\right)^2  - \gamma \right \}\ N \ ,}
where $G(ml)$ is the diagonal element of 
the lattice kernel $G({\bf x}-{\bf y})$
representing the inverse of the operator $(1 - {\nabla ^2 \over m^2})$.
Clearly, this diagonal element depends on the dimensionless
parameter $(ml)$.
Strings condense when the coefficient of $N$ becomes negative.
This  condensation condition depends crucially  upon the integer 
coprimes $p$ and $q$: if $p > q$ we will have a phase in which the
$c_\mu$ form a condensate while the $b_\mu$ are in a dilute phase 
in the regime
$$\gamma < {\pi^2 G(ml) \over 2 g^2} < {p^2 \over q^2} \gamma \ .$$
If $q >p$ the opposite happens. We will concentrate on the former
case, which is by far the most interesting.

When the topological defects $c_\mu$ condense,
the Wilson surface expectation value takes the form:

\eqn\nto{\eqalign{\langle W_S \rangle &= {1\over Z_{\rm top} } 
\sum_{\{ b_\mu\} }
 {\rm exp } \left( - S_{\rm top} - W_{\rm top} -
W_0 \right) \cr  S_{\rm top} &= \sum_{{\bf
x}, \mu } \ - {\pi^2\over 2g^2} \  Q {1 \over
\nabla ^2} Q  \ ,\cr
W_0 &= \sum_{{\bf x}, \mu } - {g^2  j^2\over l^2}\  J_{\mu \nu}^{\rm
T}{1 \over \nabla ^2} J_{\mu \nu }^{\rm T} - {2 g^2  j^2\over m^2
l^2}\  \hat d_\mu J_{\mu \nu}{1 \over \nabla ^2} \hat d_\rho
J_{\rho \nu } \ ,\cr
W_{\rm top} &= \sum_{{\bf x}, \mu } - {i \pi j\over l}\  b_\mu
{\hat K_{\mu \nu \alpha} \over \nabla ^2} J_{ \nu \alpha }^{\rm T}
 \ .\cr}}
Here $Q = l \hat d_\mu b_\mu$ represent the monopoles that live at
the end-points of the $b_\mu$ strings and $J_{ \mu \nu }^{\rm T} = -
{1 \over 2}{ O_{\mu \nu \alpha \beta} \over \nabla^2} J_{\alpha
\beta}$ is the transverse projection of the $J_{\mu \nu}$ current.
In \nto\ the longitudinal degrees of freedom of $J_{\mu \nu}$ form a
generalized Coulomb gas and are completely decoupled from the
transverse degrees of freedom. The transverse contribution can be
obtained starting from a compact Kalb-Ramond action
$$S = \sum_{{\bf x}, \mu }\  {l^4\over 2g^2} \left( F_{\mu } +{\pi
\over l^2} \tilde b_{\mu } \right) ^2 - ilj  B_{\mu \nu} J_{\mu
\nu}^{\rm T}  \ ,$$
while the Coulomb gas for the boundaries can be obtained from an
action \cpb\ with $p$ = 0 (non compact QED) coupled to $\hat d_\mu
J_{\mu \nu}$. These two gauge theories for $ J_{\mu \nu}^{\rm T}$ and
$\hat d_\mu J_{\mu \nu}$ are completely independent.

The monopoles $Q$ are always in a plasma phase \pol, \orl\
(which is not in contrast with saying that the $c_\mu$ strings condense
first, because it is their condensation that frees the monopoles at the
end-points of the $b_\mu$ strings).  In order to evaluate the effects
of the condensation of monopoles, we choose the external probe as
$J_{\mu \nu} = l d_\mu J_\nu + l  K_{\mu \nu \alpha} \phi_\alpha$, with
$J_\mu$ and $\phi_\mu$ integer, so that 
$J_{\mu \nu }^{\rm T} = l  K_{\mu \nu \alpha} \phi_\alpha$. Repeating
Polyakov's calculation \pol,
we see that monopole condensation changes the transverse contribution
to
\eqn\nwo{W_0 = \sum_{{\bf x}, \mu } - 2 g^2 j^2\ \phi_\mu
\phi_\mu +... \ ;} 
$\phi_\mu$ is the dual of the volume form $\alpha_3$
that appears in the Hodge decomposition of the current $J_{\mu \nu }$
so that \nwo\ describes the volume law in the case of closed strings.
This result is correct in the limit in which the mass $\bar m$
generated by the monopole condensation \pol\ is very large.
Away from this limit, we would
have had: 
$$W_0 = \sum_{{\bf x}, \mu } {- g^2  j^2\over l^2}\  J_{\mu \nu}^T
{1 \over {\bar m^2 - \nabla ^2}} J_{\mu \nu }^T - 2 g^2  j^2
\  \phi_\mu {\bar m^2 \over {\bar m^2 - \nabla ^2}} \phi_\mu +...\ .$$ 
This term can be obtained starting from an action of type \sla, with
$p$ = 2 and the identification $J_3 = * \phi_1$, $\delta J_3 = J^T_2$.
The appearance of a massive 3-index tensor is an example of the more
general Julia-Toulouse mechanism \pin\ and can be interpreted as
string confinement. 
After condensation of topological defects, our original gauge model for
the interactions of open 1-branes can be rewritten as the sum of two
{\it completely\ decoupled} theories, one formulated in terms of a
massless vector that describes the interactions of 0-branes (particles)
and the other formulated in terms of a massive 3-index tensor that
describes the interactions of open 2-branes (membranes).

This example can be easily generalized to the case of (d+1) dimensions
and to generic $p$. The topological defects will be, in general,
objects with higher dimensions with respect to the case in the
example. For this reason, the condensation conditions will be much
more complicated. In the following we {\it assume} that there is a
phase in which the topological defects of the lower-rank tensor
condense, and show that, in this case, the result is 
analogous to the above example.

To this end we start from a generalized version of \lam\ (we consider
$l=1$ in order to simplify the notation):

\eqn\nla{\eqalign{Z(J) &= \sum _{{\{\tilde b_{\mu_1...\mu_{d-p-1} }\},
\{\tilde a_{\mu_1...\mu_{p+1} }\} }\atop {  \{ m_{\mu_1...\mu_{p+1} }\}
}} \int_{-\pi }^{\pi }
 {\cal D} A_{\mu_1...\mu_p } {\cal D} B_{\mu_1...\mu_{p+1} } \ {\rm
exp}(-S) \ ,\cr S &= \sum_{{\bf x}, \mu }\  {1\over 4 (d-1)! g^2}
\left( F_{\mu_1...\mu_{d-p-1} } +2 \pi
 \tilde b_{\mu_1...\mu_{d-p-1} }
\right)^2 +\cr &+ {1 \over 4} \tilde m^2
 \left( B_{\mu_1...\mu_{p+1}} + 2\pi
m_{\mu_1...\mu_{p+1}} \right)^2 
 + {1\over 4e^2} 
\left( f_{\mu_1...\mu_{p+1}} + 2\pi  
\tilde a_{\mu_1...\mu_{p+1}} \right)^2 
+\cr &+ {1 \over 2e} \tilde m \left( B_{\mu_1...\mu_{p+1}} +2\pi
 m_{\mu_1...\mu_{p+1}}
 \right) \left( f_{\mu_1...\mu_{p+1}} + 2\pi
 \tilde a_{\mu_1...\mu_{p+1}} 
\right)^2 +\cr &- ij \left( B_{\mu_1...\mu_{p+1}} + { 1\over \tilde m
e}  f_{\mu_1...\mu_{p+1}} \right) J_{\mu_1...\mu_{p+1}} \ ,\cr}}
where $e$ and $g$ are still dimensionless coupling constants.
$F_{\mu_1...\mu_{d-p-1}} =
K_{\mu_1...\mu_d}B_{\mu_{d-p}}...B_{\mu_d}$ is the generalization of
the dual of the higher-rank tensor field strength, with
$K_{\mu_1...\mu_d} = S_{\mu_1} \epsilon_{\mu_1 \nu \mu_2...\mu_d}
d_\nu$  ($ \hat K_{\mu_1...\mu_d} =  \epsilon_{\mu_1...\nu \mu_d}
\hat d_\nu \hat S_{\mu_d}$).
The $K$ operators satisfy the equation:

\eqn\nke{\eqalign{&\hat K_{\mu_1...\mu_r \alpha_{r+1}...\alpha_d}
K_{\alpha_{r+1}...\alpha_d \nu_1...\nu_r}
 =  K_{\mu_1...\mu_r \alpha_{r+1}...\alpha_d} 
\hat K_{\alpha_{r+1}...\alpha_d \nu_1...\nu_r} = 
O_{\mu_1...\mu_r\nu_1...\nu_r} =\cr &=  (d-r-2)!(r+1)! \left[ -
\delta_{\mu_1 \nu_1}...\delta_{\mu_r \nu_r} \nabla^2 + (r+1) 
d_{\mu_1} \hat d_{\nu_1} \delta_{\mu_2 \nu_2}...\delta_{{\mu_r}
\nu_r}\right] \ .\cr }}
With a generalization of \sho\ we extend the integrals from $[-\pi,
\pi]$ to $(-\infty, +\infty)$ and, performing the integral over
$A_{\mu_1...\mu_p}$, we obtain:

\eqn\ebt{\eqalign{&Z(J) = \sum _{{\{ b_{\mu_1...\mu_{d-p-1} }\},
\{ c_{\mu_1...\mu_{d-p-1} }\} }} \int_{-\infty}^{\infty}
  {\cal D} B_{\mu_1...\mu_{p+1} } \ {\rm
exp}(-S) \ ,\cr &S = \sum_{{\bf x}, \mu }\  {1\over 4 (d-1)! g^2}
\left( F_{\mu_1...\mu_{d-p-1} } +2 \pi 
 b_{\mu_1...\mu_{d-p-1} } \right) ^2 - ij 
 B_{\mu_1...\mu_{p+1}} J_{ \mu_1...\mu_{p+1}}\cr &-
 {m^2 \over 4 g^2}
B_{\mu_1...\mu_{p+1}} {O_{\mu_1...\mu_{p+1}\nu_1...\nu_{p+1}} \over
\nabla^2} B_{\nu_1...\nu_{p+1}} - {\pi^2 \over e^2}
c_{\mu_1...\mu_{d-p-1}} {1 \over
\nabla^2} c_{\nu_1...\nu_{d-p-1}} + \cr &-
 {m  \over ge} B_{\mu_1...\mu_{p+1}}
{K_{\mu_1...\mu_{p+1}\nu_{p+2}...\nu_d} \over \nabla^2}
c_{\nu_{p+2}...\nu_d} - ij \hat d_\alpha B_{\alpha
\mu_2...\mu_{p+1}} {1 \over \nabla^2} \hat d_\beta J_{\beta
\mu_2...\mu_{p+1}} +\cr &+ {j^2 (p+1) g^2 \over  m^2 e^2}
\hat d_\alpha J_{\alpha \mu_2...\mu_{p+1}} {1 \over \nabla^2} \hat
d_\beta J_{\beta \mu_2...\mu_{p+1}} \ .\cr}}
The terms that describe the coupling of $B_{p+1}$ to $J_{p+1}$
can be written as

\eqn\bjc{\eqalign{ &+ ij B_{\mu_1...\mu_{p+1}} J_{
\mu_1...\mu_{p+1}} + ij \hat d_\alpha B_{\alpha
\mu_2...\mu_{p+1}} {1 \over \nabla^2} \hat d_\beta J_{\beta
\mu_2...\mu_{p+1}} = \cr
&= -{ij\over (d-p-1)!(p+1)} B_{\mu_1...\mu_{p+1}}
{O_{\mu_1...\mu_{p+1}\nu_1...\nu_{p+1}} \over \nabla^2}
J_{\nu_1...\nu_{p+1}} = ij B_{\mu_1...\mu_{p+1}}
J^T_{\mu_1...\mu_{p+1}}\ .\cr}}
$-{1\over (d-p-1)!(p+1)} {O_{\mu_1...\mu_{p+1}\nu_1...\nu_{p+1}} \over
\nabla^2}$ is the transverse projection operator. As we can see from
\bjc, the higher-rank tensor couples only to the transverse part of
the current $J_{p+1}$ and not to its boundary. 

The condensation of the topological excitations
$c_{\nu_1...\nu_{d-p-1}}$ leads to a term 
$$ {m^2 \over 4 g^2}
B_{\mu_1...\mu_{p+1}} {O_{\mu_1...\mu_{p+1}\nu_1...\nu_{p+1}} \over
\nabla^2} B_{\nu_1...\nu_{p+1}}\ ,$$ which  cancels the corresponding
one in \ebt. We are thus left with:

\eqn\fpf{\eqalign{&Z(J) = \sum_{\{ b_{\mu_1...\mu_{d-p-1}}\} }
\int_{-\infty}^{\infty}
  {\cal D} B_{\mu_1...\mu_{p+1} } \ {\rm
exp}(-S) \ ,\cr &S = \sum_{{\bf x}, \mu }\  {1\over 4 (d-1)! g^2}
\left( F_{\mu_1...\mu_{d-p-1} } +2 \pi 
 b_{\mu_1...\mu_{d-p-1} } \right) ^2 +\cr &- ij 
 B_{\mu_1...\mu_{p+1}} J^T_{ \mu_1...\mu_{p+1}} + {j^2 (p+1) g^2 \over 
m^2 e^2} \hat d_\alpha J_{\alpha \mu_2...\mu_{p+1}} {1 \over \nabla^2}
\hat d_\beta J_{\beta \mu_2...\mu_{p+1}} \ .\cr}}
The action \fpf\ describes the  theory for a compact
antisymmetric tensor of rank $(p+1)$ coupled to a conserved current 
$J^T_{ \mu_1...\mu_{p+1}}$. The last term is a generalized Coulomb gas
for the boundaries of $J_{ \mu_1...\mu_{p+1}}$.
Except for the last term, \fpf\ is exactly the theory studied in
\orl. When the system is disordered, with the same steps as in
the example, we can rewrite the theory as the sum of two
decoupled non-compact theories: one massless for closed $(p-1)$-branes
and one massive for open $(p+1)$-branes.

The generalization to the model \dba\ is easy at this point. The
lattice partition function is

\eqn\nld{\eqalign{&Z(J) = \sum_{{\{\tilde b_{\mu_1...\mu_{d-p-1} }\},
\{\tilde a_{\mu_1...\mu_{p+1} }\} }\atop {  \{ m_{\mu_1...\mu_{p+1} }\}
}} \int_{-\pi }^{\pi }
 {\cal D} A_{\mu_1...\mu_p } {\cal D} B_{\mu_1...\mu_{p+1} } \ {\rm
exp}(-S) \ ,\cr &S = \sum^M_{{\bf x}, \mu }\  {1\over 4 (d-1)! g^2}
\left( F_{\mu_1...\mu_{d-p-1} } +2 \pi
 \tilde b_{\mu_1...\mu_{d-p-1} }
\right)^2  - ij B_{\mu_1...\mu_{p+1}}  
J_{\mu_1...\mu_{p+1}} +\cr 
&+\sum^{\Sigma}_{{\bf x}, \mu } {1 \over 4} \tilde m^2
 \left( B_{\mu_1...\mu_{p+1}} + 2\pi
m_{\mu_1...\mu_{p+1}} \right)^2 
 + {1\over 4e^2} 
\left( f_{\mu_1...\mu_{p+1}} + 2\pi  
\tilde a_{\mu_1...\mu_{p+1}} \right)^2 
+\cr &+ {1 \over 2e} \tilde m \left( B_{\mu_1...\mu_{p+1}} +2\pi
 m_{\mu_1...\mu_{p+1}}
 \right) \left( f_{\mu_1...\mu_{p+1}} + 2\pi
 \tilde a_{\mu_1...\mu_{p+1}} 
\right)^2 +\cr &+ ij {(p+1) \over \tilde m
e}  A_{\mu_1...\mu_p} \hat d_\mu J_{\mu \mu_1...\mu_p} \ ,\cr}}
where $M$ indicates that the first sum is extended over the entire
space-time, while $\Sigma$ indicates that the sum is defined on an
$n$-brane.
As we said before, we must have $(d+1) \ge n \ge p$.

The topological defects associated with the lower-rank tensor live
now on the $n$-brane. With the same analysis as before, when they
condense we obtain:  

\eqn\fpd{\eqalign{&Z(J) = \sum_{\{ b_{\mu_1...\mu_{d-p-1}}\} }
\int_{-\infty}^{\infty}
  {\cal D} B_{\mu_1...\mu_{p+1} } \ {\rm
exp}(-S) \ ,\cr &S = \sum^M_{{\bf x}, \mu }\  {1\over 4 (d-1)! g^2}
\left( F_{\mu_1...\mu_{d-p-1} } +2 \pi 
 b_{\mu_1...\mu_{d-p-1} } \right) ^2 - ij 
 B_{\mu_1...\mu_{p+1}} J^T_{ \mu_1...\mu_{p+1}} +\cr 
&+\sum^{\Sigma}_{{\bf x}, \mu }\   {j^2 (p+1) g^2
\over  m^2 e^2} \hat d_\alpha J_{\alpha \mu_2...\mu_{p+1}} {1 \over
\nabla^2} \hat d_\beta J_{\beta \mu_2...\mu_{p+1}} \ .\cr}}
Again, these are the same result as in \fpf, the only
difference being that now the theory for the boundaries
is constrained to live on the $n$-brane.
When the topological defects in \fpd\ condense we will have a theory
for open $(p+1)$-branes defined in all space-time $M$, and a theory
for closed $(p-1)$-branes on the $n$-brane.

\bigbreak\bigskip\bigskip
\noindent {\bf Acknowledgements} 

\noindent We acknowledge helpful discussions with A. Giveon, S.J. Rey
and C.A. Trugenberger.

\listrefs
\end